\magnification=\magstep1
\hfuzz=3pt
\headline={\ifnum\pageno=1 \hfil\else\hss{\tenrm\folio}\hss\fi}
\footline={\hfil}
\hoffset=-.2cm
\font\titlefont= cmbx10 scaled \magstep1
\def\integer{\mathchoice{\bf N}{\bf N}{\bf N}{\bf N}}
\def\real{\mathchoice{\bf R}{\bf R}{\bf R}{\bf R}}

\def\complex{\mathchoice{\bf C}{\bf C}{\bf C}{\bf C}}

\def\Romannumeral#1{\uppercase\expandafter{\romannumeral#1}}
\def\date{\line{\number\day/\number\month/\number\year\hfil}}

\def\zinteger{\mathchoice{\bf Z}{\bf Z}{\bf Z}{\bf Z}}
\def\Oun{{\cal O}(1)}

\def\proof{\noindent{\bf Proof. }}

\def\qed{}
\def\ref#1{$^{\hbox{\sevenrm(#1)}}$}

\def\und{\underline}
\vglue 3cm
\centerline{\titlefont AMPLITUDE EQUATION FOR LATTICE MAPS,}
\vskip 1cm 
\centerline{\titlefont A RENORMALIZATION GROUP APPROACH}
\vskip 3cm
\centerline{P.COLLET}
\bigskip
\centerline{Centre de Physique Th\'eorique}
\centerline{Laboratoire CNRS UPR 14}
\centerline{Ecole Polytechnique}
\centerline{F-91128 Palaiseau Cedex (France)}
\vskip 3cm
\noindent{\bf Abstract}: 
We consider the development of instabilities of  homogeneous
stationary solutions of discrete time lattice maps. Under some generic
hypothesis we derive an amplitude equation which is the space-time
continuous Ginzburg-Landau equation. Using dynamical renormalization
group methods we control the accuracy of this approximation in a large 
ball of its basin of attraction. 
\bigskip
\noindent{\bf Keywords}: multi scales, Ginzburg Landau, bifurcations,
instabilities, continuous spectrum. 
\vfill\supereject
\beginsection{I.Introduction.}

Instabilities of extended systems lead to a large number of interesting
phenomena and in particular to the appearance of well defined 
structures\ref{5,12}. We will consider in this paper
the case of discrete time iteration of  lattice systems. 
These are dynamical systems defined on the phase  
$l^\infty(\zinteger^d,\real)$ which will be abbreviated  below by $l^\infty$.
The time evolution is given by a map $\Phi$ of $l^\infty$ into itself
for which we will make the following basic hypothesis which also
includes the dependence on a real parameter. 

\noindent{\bf H1} {\sl $\Phi\;:\; \real\times l^\infty \longrightarrow
l^\infty$. $\Phi$ is $C^2$ in its first argument and $C^4$ in the second
one on some neighborhood of the origin.
 Moreover $\Phi$ commutes with the translations on $\zinteger^{d}$.}

The time evolution for a fixed parameter $\eta$ is given as follows. If
$\underline u^t\in l^\infty$ ($t\in\integer$)  is the state of the 
system at time 
$t$, the state of the system at time $t+1$ is given by
$$
\underline u^{t+1}=\Phi(\eta,\underline u^t)\;.
$$

There are many examples of such dynamical systems. We will just mention
the class of coupled lattice maps and also notice that many 
algorithms used for solving numerically partial differential equations
are of this form (for example finite difference schemes). 
 
One would like of course to understand the properties of this dynamical
system and in particular the large time dynamics. In finite dimensional
dynamical systems, varying  a parameter is a very efficient way
to discuss the increase of complexity of the time asymptotic
dynamics. This increase of complexity can occur at well defined
threshold values of the parameter called bifurcations for which
detailed results are available in particular for the bifurcation of
stationary solutions\ref{8}. An important
tool in bifurcation theory of stationary solutions 
is the normal form equation which provides a
model for the dynamics. Moreover for generic systems these normal forms 
 are in some sense universal.

For spatially extended continuous 
systems  which depend on a parameter, amplitude
equations have been derived in various situations\ref{5,12}, and 
rigorous derivations of the
Ginzburg-Landau equation in one space dimension
 have been obtained\ref{3,15,11,13,14}.
 The result is similar to what occurs for normal forms in
finite dimensional dynamical systems, namely the space-time function
reconstructed using the solution of the amplitude equation stays close
to the solution of the true evolution (with the same initial condition) 
for a large time. However this
method usually provides only an approximation of the true solutions and
for very large times, neglected small errors usually grow due to
instabilities\ref{10}.
Nevertheless,  just before the approximation becomes too
inaccurate one can restart the process with a new initial condition
which is the present true solution. This provides a shadowing of the
true solution by a sequence of chunks of solutions of the 
amplitude equation. We refer to the previously mentioned references for
more detailed discussions of these results.

The shadowing  of a solution
of the true equation by  a reconstructed  function from a solution of
the amplitude equation is only valid for initial conditions of a rather
limited form, namely those which can be reconstructed from an initial
condition of the amplitude equation. Eckhaus\ref{6} has shown that this
condition can be somewhat relaxed and that after a relatively short time
any small enough initial condition leads to a solution having the right
form.

We will consider below all the previously mentioned results for the case
of discrete dynamics of one dimensional
 lattice systems. We will recover all the results
obtained in the continuous case and extend them in various
directions. In order to do this we will use a renormalization group
approach\ref{1,7}.
Besides its systematic simplicity this method will also have
the advantage of providing a rigorous basis for the universality of the
amplitude equations. In  the finite dimensional situation 
transversality arguments can be used, however in the infinite dimensional
case the equivalent results are not known, and universality provided by
the renormalization group method is a good replacement.

 Another  advantage of the renormalization group
argument is that it will allow us to deal with initial conditions which
are small but of order unity, independently of the smallness of the
parameter.  As we will see below the renormalization group
flow draws an initial data of size of order one to a configuration of
size $\sqrt\eta$. At the same time the renormalization group flow puts
the solution in a resonant form providing a natural proof of the
previously mentioned Eckhaus result. Finally when the
renormalization group stops improving the size and form of the solution
we get the amplitude equation together with a rigorous proof of the
associated shadowing properties. The same ideas can probably be applied
in the continuous space and time case and will eventually provide some
improvements in term of simplicity and naturalness of proof as well as
better bounds. Different approaches to amplitude equations
using renormalization have been discussed by Goldenfeld et al.\ref{4}.
 However since there is already a large literature for the
continuous case we will not consider this situation. Note also that we
will treat  a rather large class of systems whose continuous
equivalent would include pseudo differential operators.

We will mostly deal below with one dimensional systems ($d=1$) although
the same method works in higher dimensions, at least when the
amplitude equation is not too involved. Systems of coupled equations
(corresponding to vector valued fields) can be tee treated similarly.
   
Since we will be interested in instabilities of stationary solutions, 
we will now assume for simplicity
that the zero element of $l^\infty$ is a stationary
solution of the dynamical system. In many cases this can be
realized by a translation in function space.

\noindent{\bf H2} {\sl  $\Phi(\eta,\und0)=\und0$ for any $\eta\in\real$.}

 We will 
 make some assumptions on the
linearized evolution around this fixed point.

\noindent{\bf H3} {\sl We will assume that
 the operator $L_{0}=D_2\Phi(0,0)$ which is bounded in $l^{\infty}$ and
 commutes with translations can be
represented as the convolution with a (real)  sequence $l\in l^{1}$. 
Let $l(\theta)$ be the Fourier transform of $l$
$$
l(\theta)=\sum_{n\in\zinteger}l_ne^{-in\theta}\;.
$$ 
We will also assume that
$$
\sum_{n\in\zinteger}|n|^{10}|l_n|<\infty\;,
$$
and that $|l(\theta)|\le1$ with $|l(\omega)|=1$ for a unique
$\omega\in ]0,\pi[$. We will also assume that $l(\omega)\neq -1$ and
$\omega\notin\{\pi/2,3\pi/3\}$. 
 Moreover, we will make the non degeneracy assumption
$$
D=-\Re\left[{l''(\omega)\over l(\omega)}-{{l'(\omega)}^{2}\over
l(\omega)^{2}}\right] 
\neq0\;.
$$}

\noindent
As we will see later on, this last number is always non-negative.
Notice that $l(\omega)=\overline{l(-\omega)}$. The particular cases
which are excluded in the above hypothesis correspond to some strong
resonances. We will not consider them further below although some
interesting results can probably be derived in these situations using
the ideas of the present paper.

We now need some hypothesis on the higher order derivatives of $\Phi$. 

\noindent{\bf H4} {\sl The translation invariant bilinear and trilinear 
operators $Q=D_2^{2}\Phi(0,0)/2$ and $C=D_2^{3}\Phi(0,0)/6$
 are  given by a double (respectively triple) real sequence
$(q_{n,m})\in\l^{1}(\zinteger^{2})$ (respectively $(c_{n,m,p})
\in\l^{1}(\zinteger^{3})$),  namely
$$
Q(\underline u)_r=\sum_{(n,m)\in\zinteger} q_{r-n,r-m}u_n u_m
\qquad\hbox{\rm and }\qquad C(\underline u)_r=\sum_{(n,m,p)\in\zinteger}
 c_{r-n,r-m,r-p}u_n u_m u_p\;.
$$
We will assume that these sequences satisfy 
$$
\sum_{(n,m)\in\zinteger} |q_{n,m}|(1+|n|+|m|)^{6}+
\sum_{(n,m,p)\in\zinteger} |c_{n,m,p}|(1+|n|+|m|+|p|)^{4}<\infty\;.
$$
We will also make the non linear stability assumption
$$
g=-\Re\left({c(\omega,\omega,-\omega)\over
l(\omega)}+{4q(\omega,0)q(\omega,-\omega)\over
3l(\omega)(|l(\omega)|^{2}-l(0))}+
{2q(-\omega,2\omega)q(\omega,\omega)\over
3l(\omega)(l(\omega)^{2}-l(2\omega))}
\right)>0\;,
$$
where
$$
c(\phi_{1},\phi_{2},\phi_{3})=\sum_{p,q,r}c_{p,q,r}
e^{-i(p\phi_{1}+q\phi_{2}+r\phi_{3})}\qquad{\rm and}\qquad
q(\phi_{1},\phi_{2})=\sum_{p,r}q_{p,r}
e^{-i(p\phi_{1}+r\phi_{2})}\;.
$$}

Finally we will also require that the dependence on the parameter
allows the instability to occur.

\noindent{\bf H5.} {\sl  We will assume that
 the operator $L_{1}=D_1D_2\Phi(0,0)$ which is bounded in $l^{\infty}$ and
 commutes with translations can be
represented as the convolution with a (real)  sequence $m\in l^{1}$. 
We will assume that 
$$
\sum_{n\in\zinteger}|n|^{3}|m_{n}|<\infty\;.
$$
Let $m(\theta)$ be the Fourier transform of $m$, we will also assume that
$$
\rho=\Re{m(\omega)\over l(\omega)}>0\;.
$$}

\noindent Note that as for the function $l$ we have
$m(-\theta)=\overline{m(\theta)}$.

A simple example satisfying the above hypothesis can be satisfied  as
follows. Let $l$ be a bi-infinite sequence of real numbers satisfying
{\bf H3}, we define the map $\Phi$ by
$$
\Phi(\eta,\und u)_n=\eta u_n+\sum_q l_{n-q}u_q-(u_n)^3\;.\eqno (I.1)
$$
In this particular situation, {\bf H1} and {\bf H2} are easily verified. 

This is a natural generalization of the discrete Swift-Hohenberg
evolution which is given by
$$
u^{t+1}_n=(1+\eta)u^t_n-\rho\bigl((1+\gamma\Delta)^2\und u^t\bigr)_n
-(u_n^t)^3\eqno \hbox{\rm (DSH)}
$$
where $\Delta$ is the discrete Laplacian
$$
(\Delta \und u)_n=u_{n+1}+u_{n-1}-2u_{n}\;.
$$
In order to explain some of the arguments in a simpler situation we will
also sometimes make the following hypothesis.

\noindent{\bf H3'} {\sl $l$ is even and  satisfies {\bf H3}.}

\noindent Note that in this case $l(\theta)$ is even and real,
$l(\pm\omega)=1$ and $l'(\pm\omega)=0$.

For the particular case of (DSH) one has 
$$
l(\phi)=1-\rho(1-2\gamma+2\gamma\cos\phi)^2\;,
$$
and the hypothesis {\bf H1-H5} and {\bf H3'} are satisfied if one assumes
$$
\gamma>1/4\qquad\hbox{\rm and}\qquad 2/\rho>\max(1,(1-4\gamma)^2)\;.
$$

We now formulate the main result of this paper for the evolution (I.1).

\proclaim{Theorem I.1}. {Assume {\bf H3'}, then 
there are constants $R>0$, $1>\eta_{0}>0$,
$c_{1}>0,\cdots ,c_{4}>0$ such that for any $\eta\in]0,\eta_{0}[$, for
any $\underline u\in l^{\infty}$ such that $\|\underline
u\|_{l^\infty}<R$, if  $(\underline
u^{t})_{t\in\integer}$  denotes 
 the orbit of $\underline u$ under the discrete
time evolution (I.1), then
\item{i)} there is a positive number $T=T(\underline u,\eta)$ such that
for any integer $t>T$ 
$$
\|\underline u^{t}\|_{l^\infty}\le c_{1}\eta^{1/2}\;.
$$
\item{ii)} For any integer $t>T$ there is a function $A(\tau,x)$ from
$\real^{+}\times \real$ to $\complex$ satisfying
$$
\partial_{\tau}A=\partial_{x}^{2}A+A-A|A|^{2}\eqno(GL)
$$ 
such that for any integer $s$ with $0\le s\le
c_{2}\eta^{-1}\log\eta^{-1}$ we have
$$
\|\underline u^{t+s}-\underline v^{s}\|_{l^\infty}\le c_{3}\eta^{1/2+c_4}\;.
$$
where
$$
v^{s}_{n}=3^{-1/2}e^{i\omega n}\eta^{1/2}A(\eta s,\eta^{1/2}n
 \sqrt{2/D})+c.c.
$$}

Theorem (I.1) can be understood in the light of the basic principles of
bifurcation theory. First of all we have an "unstable" direction in
phase space which is the sequence $(e^{i\omega n})$ (and its complex
conjugate). As in finite dimensional bifurcation theory, the non trivial
solutions are sought in the form $u_{n}=A_{n} e^{i\omega n}$, where $A$
is a slowly varying function of space and time (it is a non trivial
consequence of our theorem that locally there are no other long time
solutions). The equation for $A$ should ensure that in the evolution
equation for $\underline u$ all resonant terms (i.e. terms containing
$e^{\pm i\omega n}$) disappear. These resonant terms come of course from
the linear part of the evolution equation but also from the non linear
part. If we examine more carefully the second contribution, we see that
quadratic terms in $\underline u$ (and all even higher order
terms) should  not contribute, whereas cubic terms (and higher odd
powers) generate resonant terms. If the solution is small, cubic terms
dominate. If we assume that the dispersion relation is non degenerate
at the critical wave number (here $D>0$), one is naturally lead by
scaling arguments to the equation
$$
\partial_{t}A=A''-A|A|^{2}\;.\eqno(F.P.)
$$  
Note that this equation has the one parameter family of scale invariance
$x\to\lambda x$, $t\to\lambda^{2}t$, $A\to\lambda A$. This equation can
be considered as a fixed point of a renormalization operation. An
important property of (F.P.) is that all solutions with bounded initial
conditions tend to zero when $t$ tends to infinity. A term $\sigma A$
added to (F.P.) corresponds to a relevant direction\ref{1}. This
is how the complete Ginzburg-Landau equation appears. The
proof of Theorem (I.1) is essentially a control of the R.G. flow in the
vicinity of (F.P.) done from the point of view of the solutions
i.e. proving that the unstable manifold is transversally stable.

We implement these ideas in  section II. We first give estimates on various
operators which are needed to control the time evolution of the
remainder. The recursive renormalization argument is given at the end of
the section. 

In section III we will give the
equivalent result under the more general hypothesis {\bf H1-H5}. 
Technically the proof  is not much different (except for the treatment
of the quadratic term). However one has to deal simultaneously 
with several details which somewhat obscure the main ideas, this is why
we have  chosen to explain first in section II the simple case, and
explain in section III the necessary modifications for the general case.

\beginsection{II. Renormalization proof of Theorem II.1.}

\def\cala{{\cal A}}
\def\calr{{\cal R}}

In this section we will prove Theorem I.1. We will therefore always
assume that $\eta$ is a positive number.

We start by fixing a renormalization factor $S$ which is a positive
number larger than one. Although the final result will
be essentially independent of $S$, it is convenient in the present
discrete space-time  setting to take for $S$ and integer (for example $S=2$).
This avoids using integer parts of various time scales.
 We can now define a sequence of
scales $(\delta_{n})_{n\in\integer}$ for the wave numbers  by
$$
\delta_{n}=S^{-n}\;.
$$ 
In other words we have a 
sequence of space scales given by $(\delta_{n}^{-1})_{n\in\integer}$.
As we will see below, since we have a "trivial" scaling, this induces a
 a sequence of time scales $(\delta_{n}^{-2})_{n\in\integer}$,
and a sequence of field scales given by $(\delta_{n})_{n\in\integer}$ . We
will also need below a sequence of scales $(\theta_{n})_{n\in\integer}$ 
 for a window of unstable wave numbers. It is convenient to choose this
sequence of the form
$$
\theta_{n}=(\delta^{-1}_{n})^{\varsigma}
$$ 
where $\varsigma$ is a positive number smaller than one and small enough. We
will for definiteness choose $\varsigma=1/5$.

As usual with renormalization group arguments, all the steps are similar
except for a change of scales. Therefore we will start by assuming that a
scale has been fixed (i.e. an $n$ in the previous sequences). Hence we
 have  a scale $\delta=\delta_{n}$  for the solution, namely we will assume from
now on  that
the initial condition $\und u$ satisfies
$$
S^{-1}\delta<\|\und u\|_{l^\infty}\le\delta\;.
$$
We also have a scale
$\theta$ for the normalized width of the window of
wave numbers. One should think of $\delta$
as a small number (but possibly of order one, i.e. a negative power of
$S$), and $\theta$ as a large number such that 
$\delta \theta^{4}$ is small (i.e. $\theta=\delta^{1/5}$).
Explicit constraints on $\delta$ and $\theta$
will be given later on. In particular we will always assume that
$\delta\theta<1/8$ (see appendix B).  

We now  introduce three operators
whose sum is the identity and which are essentially projections on
sequences  with wave numbers near  $\omega$. They are constructed as
follows. First we consider a cut-off function $\psi$ which is infinitely
differentiable, non negative with support in $[-2,2]$ and which is equal
to one on the interval $[-1,1]$. We then define a sequence  $K_{\delta,
\theta}$ 
by
$$
K_{\delta,\theta}(n)=\delta\theta
{\cal F}\psi\left(n\delta\theta\right)\;,
$$
where ${\cal F}$ denotes Fourier transform on the real line.
Convolution by $K_{\delta,\theta}$ localizes the wave numbers in an
interval of 
length $4\delta\theta$ centered at the origin. In particular, 
for $\delta\theta<\pi/2$, since that $\psi$ has compact support
we have for $-\pi\le \phi\le \pi$
$$
\sum_{n}K_{\delta,\theta}(n)e^{in\phi}=\psi(\theta^{-1}
\delta^{-1}\phi)\;.
$$
We now define
three operators $P_{\delta,\theta}^{+}$, $P_{\delta,\theta}^{-}$ and
$R_{\delta,\theta}$ by their kernels, namely 
$$
P_{\delta,\theta}^{+}(n,m)=e^{i\omega
n}K_{\delta,\theta}(n-m)e^{-i\omega m}\;, 
$$
$$
P_{\delta,\theta}^{-}(n,m)=e^{-i\omega
n}K_{\delta,\theta}(n-m)e^{i\omega m}\;, 
$$
and
$$
R_{\delta,\theta}=I-P_{\delta,\theta}^{+}-P_{\delta,\theta}^{-}\;.
$$
From the above remark about the action of the convolution by
$K_{\delta,\theta}$, we see that $P_{\delta,\theta}^{+}$ localizes the
wave numbers around 
$\omega$ and $P_{\delta,\theta}^{-}$ around $-\omega$. In particular, if
$\delta$ is 
small enough, we obtain two disjoint intervals of wave numbers.

We now define a constant $C_{1}$ by
$$
C_{1}=2\left(1+\sup_{0\le j\le 5}\|\partial^{j}
{\cal F}\psi\|_{1}\right)\;.
$$
This constant will be a convenient bound for several quantities below.

\proclaim{Lemma II.1}. {For any
$0<\delta\theta<\pi/2$, the operators $P_{\delta,\theta}^{+}$,
$P_{\delta,\theta}^{-}$ and $R_{\delta,\theta}$ have a norm 
bounded by $C_{1}$ in $l^{\infty}$.} 

\noindent 
The proof is an immediate consequence of the fact that the function
$\psi$ belongs to the Schwarz space ${\cal S}$. \qed

We now decompose a sequence $\underline u
\in l^{\infty}$ using the above three
operators. 

\proclaim{Lemma II.2}. {For any real sequence $\underline u
\in l^{\infty}$, there is
a function $B\in C^{4}(\real,\complex)$ such that
$$
e^{-i\omega n}P_{\delta,\theta}^{+}(\underline u)_{n}=B(\delta n)\quad\forall
n\in\zinteger\;, 
$$
and  satisfying
$$
\max\left\{\theta^{-j}
\|\partial^{j}B\|_{l^\infty}\;;j=0,\cdots,4\right\}
\le C_{1}\|u\|_{l^{\infty}}\;.
$$
We have also
$$
e^{i\omega n}P_{\delta,\theta}^{-}(\underline u)_{n}=\overline B(\delta
n)\quad\forall n\in\zinteger\;. 
$$}

This is an immediate consequence of the definition of
$P_{\delta,\theta}^{\pm}$ and 
interpolation. In order to fix entirely $B$ one can  use the following
explicit choice
$$
B(x)=\delta\theta
\sum_{m}{\cal F}\psi\left(x\theta-m\delta\theta\right)
e^{-im\omega}u_{m}\;.
$$
\qed

We also define the remainder $\underline r^{0}$ by
$$
\underline r^{0}=R_{\delta,\theta}\underline u\;.
$$

We now denote by $A$ the solution of the Ginzburg Landau equation
$$
\partial_{\tau}A=A''+\eta\delta^{-2}A-A|A|^{2}\;, \eqno (II.1)
$$
with initial data
$$
A(0,x)=\sqrt3 \;\delta^{-1}B(x\sqrt{D/2})\;.\eqno(II.1')
$$

It will be convenient in the sequel to denote by $\underline \cala^{t}$
 the sequence
$$
\cala^{t}_{n}=3^{-1/2}e^{in\omega}A(\delta^{2}t,n\delta
\sqrt{2/D})+c.c.\eqno (II.2) 
$$

Finally, the equation for the evolution of the remainder term 
$$
\und r^{t}=\und u^t-\delta\und \cala^t
$$ 
is given by
$$
r^{t+1}_{n}=(L\und r^{t})_{n}
-3\delta^{2}r^{t}_{n}(\cala^{t}_{n})^{2}
-3(r^{t}_{n})^{2}\cala^{t}_{n}-(r^{t}_{n})^{3}
+{\cal R}^{t}_{n}\;,\eqno (II.3)
$$
where we have denoted by $L$ the operator
$$
L\und v=l*\und v+\eta \und v\;,
$$
and  the sequence $ \und{\cal R}^{t}$ is given by
$$
{\cal R}^{t}_{n}=\delta(L\und \cala^{t})_{n}
-\delta^{3}(\cala^{t}_{n})^{3}-\delta\cala^{t+1}_{n}\;.\eqno (II.4)
$$

Using equation (II.1), we have immediately the following estimate for
the forcing term $\und{\cal R}^{t}$. 

\proclaim{Lemma II.3}. {There is a constant $C_{2}>C_{1}$ such that for any 
$\delta\theta^{4}<1$, and $\delta>\eta^{1/2}$, 
 if $B$ is a function of $x$ satisfying
$$
\max\left\{\theta^{-j}
\|\partial^{j}B\|_{l^\infty}\;;j=0,\cdots,4\right\}
\le \delta C_{1}\;,
$$ 
if $A$
denotes the solution of (II.1) with initial condition (II.1'),  and if
$\und\cala^{t}$ is 
constructed from $A$ as in (II.2), then  for any $t\in\integer$ we have 
$$
\|\und \cala^{t}\|_{l^{\infty}}\le C_{2}\;,\;
\|\und{\cal R}^{t}\|_{l^{\infty}}\le 
C_{2}\delta^{3}\;,\;\;
\|\und{\cal R}^{t}+\und{\cal C}^{t}
\|_{l^{\infty}}\le C_{2}\delta^{4}\theta^{4}\;,\;\;
$$
where the sequence $\und{\cal C}^{t}$ is given by
$$
{\cal C}^{t}_n=3^{-3/2}\delta^{3}e^{3in\omega}A^{3}(\delta^{2}t,n\delta
\sqrt{2/D})+c.c.
$$}

\proof It follows easily from (II.1') that
$$
\|A(0,\,\cdot\,)\|_{l^{\infty}}\le \sqrt3C_{1}\;. 
$$
The Lemma follows then at once from Lemmas A.3 and A.4 of appendix A using
$\delta\theta^{4}<1$.\qed

We now start estimating various operators needed below in the proof of
Theorem I.1.

\proclaim{Lemma II.4}. {There is a constant $C_{3}>e$ such that for any
$0<\eta\le1$ and for  any $t\in\integer$  we have
$$
\|L^{t}\|_{l^{\infty}}\le C_{3}e^{\eta(t-\eta^{-1})}\;.
$$}

\noindent The proof is given in appendix B.

\proclaim{Lemma II.5}. {For any $t\in\integer$, $0<\eta<1$
 and $1>\delta^2\theta^2>4D\eta$ we have
$$
\|L^{t}R_{\delta,\theta}\|\le C_4e^{-tD\delta^{2}\theta^{2}/8}\;.
$$}

\noindent The proof is given in appendix B. 

It will be convenient to  denote by $L_{t}$ the linear operator given by
$$
(L_{t}\und v)_{n}=-3\delta^{2}(\cala^{t}_{n})^{2}v_{n}
-3\delta r^{t}_{n}\cala^{t}_{n}v_{n}-(r^{t}_{n})^{2}v_{n}\;.
$$
Note that
$$
r^{t+1}_{n}=(L\und r^{t})_{n}+(L_{t}\und r^{t})_{n}
+{\cal R}^{t}_{n}\;.\eqno (II.5)
$$

\proclaim{Lemma II.6}. {We have for any $t>0$
$$
\|L_{t}\|_{l^\infty}
\le 5 (\delta^{2}\|\und \cala^{t}\|^{2}_{l^\infty}
+\|\und r^{t}\|^{2}_{l^\infty})\;.
$$}

\noindent This is obvious from the definition of $L_{t}$.\qed

\proclaim{Lemma II.7}. { If $0\le t_{0}<t\le\eta^{-1}$
we have
$$
\|(L+L_{t})\cdots(L+L_{t_{0}})\|_{l^\infty}\le C_{3} 
\left(1+\sup_{t_{0}\le s\le t}5C_{3}
 (\delta^{2}\|\und \cala^{s}\|^{2}_{l^\infty}
+\|\und r^{s}\|^{2}_{l^\infty})\right)^{t-t_{0}+1}\;.
$$}

\proof
We first expand the product $(L+L_{t})\cdots(L+L_{t_{0}})$. It is a
sum of $2^{t-t_{0}+1}$ terms of the form
$$
 L^{1-\zeta_{t}}L_{t}^{\zeta_{t}}\cdots L^{1-\zeta_{t_{0}}}
L_{t_{0}}^{\zeta_{t_{0}}}\;,
$$
where the numbers $\zeta_{\cdot}$ are equal to zero or one. Using Lemmas
II.4 and II.6 it easily follows that
$$
\| L^{1-\zeta_{t}}L_{t}^{\zeta_{t}}\cdots L^{1-\zeta_{t_{0}}}
L_{t_{0}}^{\zeta_{t_{0}}}\|_{l^\infty}\le 
C_{3}^{k+1}5^{k} \left(\sup_{t_{0}\le s\le t}
(\delta^{2}\|\und \cala^{s}\|^{2}_{l^\infty}
+\|\und r^{s}\|^{2}_{l^\infty})\right)^{k}
$$
where $k$ is the number of $\zeta_{\cdot}$ equal to one. Therefore
$$
\|(L+L_{t})\cdots(L+L_{t_{0}})\|_{l^\infty}\le  
\sum_{k=0}^{t-t_{0}+1}\pmatrix{t-t_{0}+1\cr k\cr}
C_{3}^{k+1}5^{k} \left(\sup_{t_{0}\le s\le t}
(\delta^{2}\|\und \cala^{s}\|^{2}_{l^\infty}
+\|\und r^{s}\|^{2}_{l^\infty})\right)^{k}\hskip -7pt\;,
$$
and the result follows.\qed

If $t-t_{0}>\eta^{-1}$, we can split this time interval in various
sub-intervals of length $[\eta^{-1}]$ plus a remainder. We immediately
obtain the following corollary. 

\proclaim{Corollary II.8}. {For any $0<t_0<t$ and $\eta<1/2$ we have
$$
\|(L+L_{t})\cdots(L+L_{t_{0}})\|_{l^\infty}
\le C_3^{1+2\eta(t-t_0)}
e^{5C_3(t-t_0+1)\sup_{t_0\le s\le t}
 (\delta^{2}\|\und \cala^{s}\|^{2}_{l^\infty}
+\|\und r^{s}\|^{2}_{l^\infty})}\;.
$$}

We now start estimating the remainder $\und r^{t}$. This will be done in
two steps. Namely, during a time of order $\delta^{-2}$ we do not loose
too much on the estimate of the remainder. After this time has elapsed
and up to a time of order ${\cal O}(1)\delta^{-2}\log\delta^{-1}$ the
estimate improves because the remainder is in a contracting subspace of
the linear evolution. When we reach this time (after which we may somewhat
loose again on the estimate), we perform a
renormalization step and start the whole process again.

\proclaim{Lemma II.9}. { There is a constant $1>C_4>0$
such that if 
$C_{4}>\delta^2>\eta$,  if  $t<C_{4}\delta^{-2}$ and
$\theta<\delta^{-1/4}/(2C_{2})$ then
$$
\|\und r^{t}\|_{l^\infty}<4C_{2}C_{3}\delta\;.
$$}

\proof
Let the non decreasing sequence of positive numbers $\sigma_{t}$
 be defined by
$$
\sigma_{t}=\delta^{-1}\sup_{0\le s\le t}\|\und r^{s}\|_{l^\infty}\;.
$$
Note that by definition of $C_{1}$ we have $\sigma_{0}\le C_{1}<4C_{2}C_{3}$.
We have from equation (II.5) the identity
$$
\und r^{t+1}=(L+L_{t})\cdots(L+L_{0})\und r^{0}
+\sum_{j=0}^{t}(L+L_{t})\cdots(L+L_{j+1})\und{\cal R}^{j}\;.
$$
Using  Lemma II.3 and Lemma II.7 we obtain
$$
\eqalign{&\|\und r^{t+1}\|_{l^\infty}\cr
&\le C_{3}\left(1+5C_{3}\delta^{2}(C_2^2+
\sigma_{t}^{2})\right)^{t+1}\|\und r^{0}\|_{l^\infty}
+\sum_{j=0}^{t}C_{3}\left(1+5C_{3}\delta^{2}(C_2^2+
\sigma_{t}^{2})\right)^{t-j}\|\und{\cal R}^{j}\|_{l^\infty}\cr
&\le C_{3}(C_{1}+C_{2})\delta e^{(t+1)\delta^{2}5C_{3}
(C_2^2+\sigma_{t}^{2})}\cr}
$$
since $t\delta^{2}<1$. The result now follows recursively
from the choice of the constant 
$$
C_{4}={\log2\over
170C_{3}^{3}C_{2}^{2}}\;.
$$  
\qed

We now improve the previous result using the support properties of the
spectrum of $\und r^{0}$ and of $\und {\cal R}^{t}$. We first derive an
equivalent of Corollary II.7.

\proclaim{Lemma II.10}. {There is a  constant $C_5>1$ such that
if $\eta<1/2$,  $\delta\theta<1$, $\delta^{2}>\eta$,  and
 $t$ is such that for $0\le s\le t$ we have
$\|\und r^{s}\|\le 4C_{2}C_{3}\delta$, then
$$
\|(L+L_{t})\cdots(L+L_{0})R_{\delta,\theta}\|_{l^\infty}\le 
C_5\theta^{-2}e^{C_5t\delta^2}
+C_{5}e^{-t\delta^{2}\theta^{2}/C_{5}}\;.
$$}

\proof We first observe that
$$
(L+L_{t})\cdots(L+L_{0})=L^{t+1}+\sum_{s=0}^{t}
(L+L_{t})\cdots(L+L_{s+1})L_{s}L^{s}\;,
$$ 
where as usual, the elements with indices out of range are equal to
identity. We now apply Lemmas II.5-II.7 and Corollary II.8 to get
$$
\|(L+L_{t})\cdots(L+L_{0})R_{\delta,\theta}\|_{l^\infty}\le 
C_4 e^{-tD\delta^{2}\theta^{2}/8}+5
(C_2^2+16C_2^2C_3^2)\delta^{2}
$$
$$
\sum_{s=0}^{t-1}e^{-sD\delta^2\theta^2/8}C_3^{1+2\eta(t-s)}
e^{5C_3(t-s)\delta^2(C_2^2+16C_2^2C_3^2)}
\;,
$$
from which the result follows immediately.\qed

\proclaim{Corollary II.11}. {There are constants $C_{6}>C_{5}$ and
$C_{7}>0$ such that if  $\delta$ is small enough so that 
the numbers $\pm 3\omega\pmod {2\pi}$ do not belong to 
$[\omega-2\theta\delta,\omega+2\theta\delta]
\cup[-\omega-2\theta\delta,-\omega+2\theta\delta]$,then 
if $t$ is such that for $0\le s\le t$ we
have $\|\und r^{s}\|\le 4C_2C_3\delta$, and if
$C_{7}\delta^{-2}\log\delta^{-1} >t\ge C_{4}\delta^{-2}$ we have
$$
\|\und r^{t}\|\le C_{6}\delta\theta^{-1/2}\;.
$$}

The proof is similar to the proof of Lemma II.9 using Lemma II.10,
Corollary C1, and the relation
$$
\sum_{j=0}^{t}(L+L_{t})\cdots(L+L_{j+1})\und{\cal R}^{j}=
$$
$$
\sum_{j=0}^{t}L^{t-j}\und{\cal R}^{j}+
\sum_{j=0}^{t}\sum_{s=j+1}^{t}
(L+L_{t})\cdots(L+L_{s+1})L_{s}L^{s-j-1}\und{\cal R}^{j}\;.
$$ \qed

We can now give the recursive proof of Theorem I.1. 
We start by selecting an integer $n_{0}$ large enough so that all the
constraints on $\delta=\delta_{n}$ and $\theta=\theta_{n}$ in the
previous Lemmas are satisfied for $n>n_{0}$. We also assume $n_{0}$
large enough so that $\log\delta_{0}^{-1}>(S+1)^{2}+1$.
We define $R=\delta_{0}$.
Note that the choice of $n_{0}$ is independent of $\eta$. We now define
$\eta_{0}$ by $\eta_{0}=S^{-4}\delta_{0}^{2}/8$. 

Since the proof is 
recursive we will therefore assume that we have reached a situation
corresponding to scale $\delta_{n}$ ($n<n_{0}$). In other words we
assume that $\underline u^{0}$ satisfies
$$
\|\underline u^{0}\|_{l^{\infty}}\le \delta_{n}\;.
$$ 
We will also assume that 
$$
\|\underline u^{0}\|_{l^{\infty}}>S^{-1} \delta_{n}\;,
$$ 
otherwise the choice of $n$ is not optimal. Finally since we are for the
moment interested in the first part of the Theorem, we will also assume
that $\delta_{n}^{2}>\eta$.

We now consider the time evolution starting from this initial data
$\underline u^{0}$, and we will apply the previous Lemmas with 
$\delta=\delta_{n}$ and $\theta=\theta_{n}=\delta_{n}^{-\varsigma}$. We first
apply Lemma II.2 to get a function $B$ with the corresponding estimates
on the derivatives, and we define the function $A(\tau,x)$
 solution of (II.1) with initial data (II.1'). We also define the remainder 
$\underline r^{0}$.

The time evolution will now be split into two parts.
We  define a first time 
$$
t_{n}=[C_{4}\delta_{n}^{-2}]\;.
$$
On the time interval $[0,t_{n}]$, we have from Lemma II.9
$$
\|\underline r^{t}\|_{l^{\infty}}\le 4C_{2}C_{3}\delta_{n}\;.
$$
We now define a second time by 
$$
T_{n}=[C_{8}\delta_{n}^{-2}\log\delta_{n}^{-1}]\;,
$$
where $C_{7}>C_{8}>0$ and $C_{8}$ is small enough so that for
$t<C_{8}\delta^{-2}\log\delta^{-1}$, all the exponentially growing
factors which appear in all the previous Lemmas are smaller than
$\theta^{1/4}$.   
\def\caltn{{\cal T}_{n}}

On the time interval $[t_{n},T_{n}]$, we can now apply Lemma II.11. In
particular, there is a time $\caltn$ which satisfies
$$
t_{n}\le\caltn<T_{n}
$$
where we have
$$
\|\underline r^{\caltn}\|_{l^{\infty}}\le S^{-1}\delta_{n}/2\;,
$$
and also using Lemma A.2
$$
\|3^{-1/2}A(\delta_{n}^{2}\caltn,\,\cdot\,)\|_{L^{\infty}}\le S^{-1}/2\;.
$$
Therefore 
$$
\|\underline u^{\caltn}\|_{l^{\infty}}\le \delta_{n+1}\;,
$$
and we can now repeat the argument with $n+1$.

The proof of the second part of Theorem I.1 is essentially similar.
 We first define an integer $N$ as the smallest integer such that 
$\delta_{N}\le S\eta^{1/2}$. We can now repeat the above proof and the
result follows since $T_{N}$ is of the right order.
\beginsection{III. The General Case.}

\def\call{{\cal L}}
In this section we will prove the following theorem which is a
generalization of Theorem I.1.

\proclaim{Theorem III.1}. {Assume hypothesis {\bf H1-H5}.
Then there exists  constants  $R>0$, $\eta_{0}>0$,
$c_{1}>0,\cdots ,c_{4}>0$ such that for any $\eta\in]0,\eta_{0}[$, for
any $\underline u\in l^{\infty}$ such that $\|\underline
u\|_{l^\infty}<R$, if $(\underline
u^{t})_{t\in\integer}$  denotes the orbit of $\underline u$ under the discrete
time evolution $\Phi$, then
\item{i)} there is a positive number $T=T(\underline u,\eta)$ such that
for any integer $t>T$ 
$$
\|\underline u^{t}\|_{l^\infty}\le c_{1}\eta^{1/2}\;.
$$
\item{ii)} Define the constants $\alpha$, $\beta$, $\sigma$, $\gamma$
and $V$  by
$$
\alpha=\Im{l''(\omega)\over Dl(\omega)}\;, \qquad
e^{i\sigma}=l(\omega)\;,\qquad \gamma=\Im{m(\omega)\over
 \rho l(\omega)}\;, \qquad V=-i{l'(\omega)\over l(\omega)}\;,
$$
and
$$
\beta=-\Im\left({c(\omega,\omega,-\omega)\over
gl(\omega)}+{4q(\omega,0)q(\omega,-\omega)\over
3gl(\omega)(|l(\omega)|^{2}-l(0))}+
{2q(-\omega,2\omega)q(\omega,\omega)\over
3gl(\omega)(l(\omega)^{2}-l(2\omega))}
\right)\;.
$$
For any integer $t>T$ there is a function $A(\tau,x)$ from
$\real^{+}\times \real$ to $\complex$ satisfying
$$
\partial_{\tau}A=(1+i\alpha)\partial_{x}^{2}A+A-(1+i\beta)A|A|^{2}\eqno(CGL)
$$ 
such that for any integer $s$ with $0\le s\le
c_{2}\eta^{-1}\log\eta^{-1}$ we have
$$
\|\underline u^{t+s}-\underline v^{s}\|_{l^\infty}\le c_{3}\eta^{1/2+c_4}\;.
$$
where
$$
v^{s}_{n}=e^{i(\sigma t+\eta\gamma t+\omega n)}\eta^{1/2}
\sqrt{{\rho\over 3g}}A\big(\rho\eta t,\eta^{1/2}\sqrt{2\rho/D}(n+Vt)\big)
+c.c.\;.
$$}

We recall that the constants $D$ and $\rho$ have been defined in
hypothesis {\bf H3} and {\bf H5} respectively.

\proclaim{Lemma III.2}. {The number $D$ is non negative, and the number
$V$ is real.}

\proof Let $\zeta(\theta)=l(\theta)/l(\omega)$, this function is equal
to one in $\theta=\omega$ and this is a point where its modulus is
 maximum. Let $\zeta_{R}$ and $\zeta_{I}$ denote the real and imaginary
parts of $\zeta$. Writing that the first derivative of the square of the
modulus of $\zeta$ in $\omega$ is equal to zero and the second
derivative is non-positive, one gets immediately (using
$\zeta_{I}(\omega)=0$)
$$
\zeta_{R}'(\omega)=0\;,\qquad\hbox{\rm and}\qquad \zeta_{R}''(\omega)+
\big(\zeta_{I}'(\omega)\big)^{2}\le0\;,
$$
and the lemma follows.\qed

From the differentiability of $\Phi$ we can write
$$
\Phi(\eta,\und u)=L'\und u+\eta L''\und u+Q(\und u,\und u)+C(\und
u,\und u, \und u)+G(\eta,u)\;,\eqno(III.1)
$$
where $G$ is a nonlinear map, differentiable which is at least of degree
4 in $\und u$, at least of degree 2 in $\eta$ and at least of degree one in
 $\eta(\und u)^{2}$. 

 Our first goal is to eliminate the quadratic term $Q$. As 
we will see, this is not entirely possible, but it will be enough to
eliminate the dominant part of this map. The main idea is of course that
$Q$ should not give rise to a resonant term, and therefore it should be
possible to eliminate the quadratic terms by a nonlinear change of
variables. An implementation of this idea is given by the following
result.

\proclaim{Theorem III.3}. {There is a (differentiable)
  quadratic operator $S$
in $l^{\infty}$ such that if ${\cal S}$ denotes the operator
$$
{\cal S}(\und u)=\und u+S(\und u)\;,
$$
then on some neighborhood of zero ${\cal S}$ is invertible (with
differentiable inverse), and  the maps
$$
\tilde\Phi(\eta,\cdot)={\cal S}^{-1}\circ \Phi(\eta,\cdot)\circ {\cal S}
$$
satisfy the hypothesis {\bf H1-H5}  with a quadratic part
$\tilde Q=D_{2}^{2}\tilde\Phi(0,0)/2$ such that its Fourier transform
$\tilde q(\phi_{1},\phi_{2})$ is identically zero on a neighborhood of
the points $(\pm\omega,\pm\omega)$ and of the lines
$\phi_{1}+\phi_{2}=\pm\omega$. $\tilde q$ satisfies also
$$
\sum_{n,m}(1+|n|+|m|)^{6}|\tilde q_{n,m}|<\infty\;.
$$ 
Moreover, we have
$$
D_{2}\tilde\Phi(0,0)=L'\;,\qquad D_{1}D_{2}\tilde\Phi(0,0)=L''\;,
$$
and 
$$
D_{2}^{3}\tilde\Phi(0,0)(\und v,\und v,\und v)/6=C(\und v,\und v,\und
v)+2Q(\und v,S(\und v))\;, 
$$
and the Fourier coefficients of this form decay also polynomially fast.}

\proof 
 By continuity, it follows from {\bf H3} that 
 we can find a number $h<\omega/4$ such that if
$\phi_1$ and $\phi_2$ belong to the intervals
$[\pm\omega-2h,\pm\omega+2h]$ then
$l(\phi_1+\phi_2)-l(\phi_1)l(\phi_2)\neq 0$. Moreover, we can choose
$h$ small enough such that if $|\phi_{1}+\phi_{2}\pm\omega|<2h$ then
either $|\phi_{1}\pm\omega|>2h$ or $|\phi_{2}\pm\omega|>2h$ and 
$l(\phi_1+\phi_2)-l(\phi_1)l(\phi_2)\neq 0$.  Let
$$
\psi_{0}(\theta)=\psi((\theta+\omega)h^{-1}/2)
+\psi((\theta-\omega)h^{-1}/2)\;,
$$
 where $\psi$ is the
cut-off function defined at the beginning of section 2.
 We now define a double
sequence $(\Gamma_{n,m})$ by
$$
\Gamma_{n,m}={1\over 4\pi^{2}}\int\int
 {\psi_{0}(\phi_{1})\psi_{0}(\phi_{2})+\psi_{0}(\phi_{1}+\phi_{2})
\over l(\phi_1+\phi_2)-l(\phi_1)l(\phi_2)}
e^{-i(\phi_1n+\phi_2m)}d\phi_1d\phi_2\;.
$$  
Since the function $\psi$ defined in section 2 is regular with compact
support, it follows that the sequence $(\Gamma_{n,m})$ belongs to
$l^{1}(\zinteger^{2})$.
Note also that this sequence is real. Moreover, we have
$$
\sum_{n,m}|\Gamma_{n,m}|(1+|n|+|m|)^{k}<\infty
$$ 
for any integer $k$.

 We now define the double sequence $s$ by
$$
s=\Gamma*q\;,
$$ 
and the quadratic map $S$ by
$$
S(\und v)_{n}=\sum_{p,q}s_{n-p,n-q}v_{p}v_{q}\;.
$$
The proof follows by easy computations.\qed

We will now drop the tilde on the various transformations, i.e. we will
 assume that the function $q(\phi_{1},\phi_{2})$ satisfies the
conclusions of the previous Theorem.  We can recover the general result
 by applying the above Theorem.

The proof of Theorem III.1 is now very analogous to the proof of Theorem
I.1. It goes along the same steps of the renormalizaztion group using
estimates analogous to those of Lemmas II.3-II.11. We will only indicate the
non-elementary changes in the proofs without  keeping a detailed
account of the various constants, i.e. the notation $\Oun$ will be used
for any constant independent of $\delta$, $\eta$ and $\und u$.

 We first decompose $\und u^{0}$ as in section
2, namely we first apply Lemma II.2 and we obtain a function $B$.
As in section II, we set
$$
\und u^{t}=\delta\und \cala^{t}+\und r^{t}
$$
where now
$$
\cala^{t}_n=
e^{i(\sigma t+\omega n)}
(3g)^{-1/2}A(\delta^{2} t,\delta\sqrt{2/D}(n+Vt))
\;,\eqno (III.2) 
$$
with $A$  a solution of the complex Ginzburg Landau equation
$$
\partial_\tau A=(1+i\alpha)A''+(\rho+i\gamma)\eta\delta^{-2}A
-(1+i\beta)A|A|^{2}\;,\eqno (CGL)
$$
with initial condition
$$
A(0,x)=\sqrt{3g}\delta^{-1}B(x\sqrt{D/2})\;.\eqno(CGL')
$$
The remainder $\und r^{t}$ satisfies the equation
$$
\und r^{t+1}=(\call_{t}+L_{t})\und r^{t}+\und {\cal R}^{t} \;,
\eqno (III.3)
$$
where 
$$
\und{\cal R}^{t}=\delta L'\und\cala^{t}+\eta\delta L''\und\cala^{t}
+\delta^{2}Q(\und\cala^{t},\und\cala^{t})+
\delta^{3}C(\und\cala^{t},\und\cala^{t},
\und\cala^{t})+G(\eta,\delta\und\cala^{t})-\delta\und\cala^{t+1}\;,
$$
and the linear operators $L_{t}$ and $\call_{t}$ are  defined by
$$
\call_{t}\und v=L'\und v+\eta L''\und v+
Q(\und v,2\delta\und\cala^{t}+\und r^{t})
$$
and
$$
L_{t}\und v=
3C(\und v,\delta\und\cala^{t},\delta\und\cala^{t}+\und r^{t})
+C(\und v,\und r^{t},\und r^{t})
+\int_{0}^{1}d\xi D_{2}G(\eta,\delta\und\cala^{t}+\xi\und r^{t})\und v\;.
$$

We first describe the estimations on the linear operators. Note that
Lemma II.7, Corollary II.8 and Lemma II.10 follow as before if we have
Lemmas II.4-II.6. 

The following equivalent of Lemma II.6 is immediate from the above formulas.

\proclaim{Lemma III.4}. {If $\|\delta\und\cala^{t}\|_{l^\infty}
+\|\und r^{t}\|_{l^\infty}<1$, then
$$
\|L_{t}\|_{l^\infty}\le\Oun(\delta^{2}\|\und\cala^{t}\|_{l^\infty}^{2}
+\|\und r^{t}\|_{l^\infty}^{2})\;.
$$}

We now come to the equivalent of Lemma II.4 and Lemma II.5.

\proclaim{Lemma III.5}. {There are two constants $c>1$ and $c'>0$
such that for 
 any $\eta<\delta^{2}\theta^{2}\in ]0,b/2]$ ($b$ is the constant of
Lemma C3) and for any $t\ge t_{0}\ge0$ such that
$$
\sup_{t_{0}\le s\le t}(\delta\|\und\cala^{t}\|_{l^\infty}
+\|\und r^{t}\|_{l^\infty})\le c'\delta\;,
$$
 we have
$$
\|\call_{t}\cdots\call_{t_{0}}\|_{l^\infty}\le c e^{c\eta (t-t_{0})}
$$
and 
$$
\|\call_{t}\cdots\call_{t_{0}}R_{\delta,\theta}\|_{l^\infty}\le
c e^{-\delta^{2}\theta^{2}(t-t_{0})/c}\;.
$$}

\proof We first observe that $\call_{t}=L+\call'_{t}$
where
$$
L\und v=L'\und v+\eta L''\und v\qquad\hbox{\rm and}\qquad
\call'_{t}=Q(\und v,2\delta\und\cala^{t}+\und r^{t})\;.
$$
With this notation we have
$$
\call_{t}\cdots\call_{t_{0}}
=L^{t-t_{0}+1}+\sum_{s=t_{0}+1}^{t}
L^{t-s}\call_{s-1}'\call_{s-2}\cdots\call_{t_{0}}\;.
$$
Using Lemma C3 and Lemma II.4 we get
$$
\|\call_{t}\cdots\call_{t_{0}}\|_{l^\infty}\le
$$
$$
\Oun e^{c\eta(t-t_{0}+1)}+\sum_{s=t_{0}+1}^{t}\Oun e^{-b(t-s)}
(\delta\|\und\cala^{s-1}\|_{l^\infty}+\|\und r^{s-1}\|_{l^\infty})
\|\call_{s-2}\cdots\call_{t_{0}}\|_{l^\infty}\;.
$$
The proof of the first part of the Lemma follows now recursively.

To get the second part, we use the following formula
$$
\call_{t}\cdots\call_{t_{0}}
=L^{t-t_{0}+1}
$$
$$
+\sum_{k=1}^{t-t_{0}+1}\;\;\sum_{n_{1}+\cdots+n_{k+1}=t-t_{0}+1-k\atop
{\scriptstyle n_{1}\ge0,\;\cdots\;,\,n_{k+1}\ge0}}
L^{n_{1}}\call'_{t-n_{1}}L^{n_{2}}\call'_{t-n_{1}-n_{2}}L^{n_{3}}
\cdots
L^{n_{k}}\call'_{t-n_{1}-\cdots-n_{k}-k}
\;L^{n_{k+1}}\;.
$$
The second part of the Lemma follows now easily using Lemma II.5 and
Lemma C3. \qed

As explained before, Lemmas II.7, Corollary II.8 and Lemma II.10 follow
as before.

We now come to the estimates concerning the forcing term
$\und\calr^{t}$. From Lemma C2, the term 
$\delta^{2}Q(\und\cala^{t},\und\cala^{t})$ 
is at least of order $\delta^{4}\theta^{2}$. For the term
 $\und\calr^{t}-\delta^{2}Q(\und\cala^{t},\und\cala^{t})$we have the
same estimates as in Lemma II.3 and therefore we get bounds analogous to
those of Lemmas II.9 and II.11.

The proof of Theorem III.1 then follows
exactly the same steps as the proof of Theorem I.1 and we will not
repeat them here.  Finally
one has to apply the inverse of the map ${\cal S}$ to conclude the proof.

\beginsection{Appendix A. Some estimates on the solutions of  CGL.}

In this appendix we will derive estimates on the solution of equation
(CGL). General estimates were derived before\ref{2}, however the present
situation is slightly better in the sense that the initial conditions
will never be larger than some predefined constant. We will also
restrict ourselves to  dimensions one and two where the complex
Ginzburg-Landau equation is known to have well behaved solutions for any
values of the parameters $\alpha$ and $\beta$.  We will in fact obtain a
stronger result than what we need in sections II and III.

We consider the equation 
$$
\partial_{t}A=(1+i\alpha)\Delta A+\sigma(\rho+i\gamma) A-
(1+i\beta) A|A|^{2}\;,\eqno (A.1)
$$
where $\sigma$ is a positive parameter smaller than a given positive
constant $S$, and $\alpha$,
$\beta$, $\gamma$ and $\rho$ are fixed constants. Note that we could
have as well rescaled $\sigma$ and fixed for example $\rho=1$. We will
not do this in order to have an immediate application of the result to
section III. 

\proclaim{Lemma A.1}. {Let $A(t,x)$ be a solution of (A.1).
Assume that there is a number $k\ge2$, 
 a constant $K>1$ and a constant $\theta>1$ such that
$$
\sup_{0\le j\le k}\theta^{-j}\|\partial_{2}^{j}
A(0,\,\cdot\,)\|\le K\;.
$$
Then there is a constant $M$ which depends only on $K$, $S$, $\alpha$,
$\beta$, $\gamma$, $\rho$, such that for any
$t>0$ 
$$
\sup_{0\le j\le k}\theta^{-j}\|\partial_{2}^{j}A(t,\,\cdot\,)
\|_{L^{\infty}}\le M\;.
$$
Moreover, for any integer $p\le k/2$ we have
$$
\sup_{0\le j\le p}\theta^{-2j}\|\partial_{1}^{j}A(t,\,\cdot\,)
\|_{L^{\infty}}\le M\;.
$$
There is a number $U>0$ which is independent of $\theta$ and $A$ such
that for $t>U$ the same estimates hold without the term $\theta^{-j}$.
We can take $U$ as small as needed eventually increasing $M$.}

\proof Let $G_{t}$ denote the function
$$
G_{t}(x)={1\over (2\pi t)^{d/2}}e^{-x^{2}/4t}\;.
$$
 We  then consider the integral equation
$$
A(t,\,\cdot\,)=e^{i\alpha t}G_{t}*A(0,\,\cdot\,)
$$
$$
+\int_{0}^{t}ds G_{t-s}*\big(\sigma(\rho+i\gamma)A(s,\,\cdot\,)
-(1+i\beta)|A(s,\,\cdot\,)|^{2}A(s,\,\cdot\,)\big)\;.\eqno (A.2)
$$
It is easy to verify that there is a number $1>T>0$ which depends only on
$k$ and $K$, $S$, $\alpha$, $\beta$, $\gamma$, $\rho$,
 such that this equation can be solved by the contraction
mapping method
on the time interval $[0,T]$ with a solution in the ball of radius $2K$
in $L^{\infty}$. Moreover, the solution is also a
solution of (A.1) with initial data $A(0,\,\cdot\,)$. Taking the
successive derivatives in $x$ of (A.2) we get
$$
\partial^{j}A(t,\,\cdot\,)=e^{i\alpha t}G_{t}*\partial^{j}A(0,\,\cdot\,)
+\int_{0}^{t} G_{t-s}*\big(\sigma(\rho+i\gamma)\partial^{j}A(s,\,\cdot\,)
$$
$$
-2(1+i\beta)|A(s,\,\cdot\,)|^{2}\partial^{j}A(s,\,\cdot\,)
-(1+i\beta)A(s,\,\cdot\,)^{2}\partial^{j}\overline A(s,\,\cdot\,)
\big)\,ds+F_{j}\;,
$$ 
where $F_{j}$ is a combination of derivatives of $A$ of lower order.
Eventually taking a smaller $T$ (independently of $j$) we can again
solve this equation for $\partial^{j}A$ by contraction and obtain recursively
estimates on the derivatives up to order $k$ with the right dependence
in $\theta$. The first part of the 
Lemma is then proven on the time interval $[0,T]$.

We now observe that at time $T$ the estimate is in fact better.
Indeed what we said before is also true for the time interval $[0,T/2]$.
We can now write another integral equation, namely for $T/2<\tau\le T$
$$
A(\tau,\,\cdot\,)=e^{i\alpha (\tau-T/2)}G_{(\tau-T/2)}*A(T/2,\,\cdot\,)
$$
$$
+\int_{T/2}^{\tau}ds \;G_{\tau-s}*\big(\sigma(\rho+i\gamma)A(s,\,\cdot\,)
-(1+i\beta)|A(s,\,\cdot\,)|^{2}A(s,\,\cdot\,)\big)\;,
$$
and for the gradient 
$$
\nabla A(\tau,\,\cdot\,)=e^{i\alpha (\tau-T/2)}
\nabla G_{(\tau-T/2)}*A(T/2,\,\cdot\,)
+\int_{T/2}^{\tau}\nabla G_{\tau-s}
*\big(\sigma(\rho+i\gamma)\partial^{j-1}A(s,\,\cdot\,)
$$
$$
-(1+i\beta)|A(s,\,\cdot\,)|^{2}A(s,\,\cdot\,)\big)\;ds\;,
$$
We then  deduce from the integrability in $t=0$ of the $L^{1}$ norm (in
$x$) of  $\nabla G_{t}$ an estimate on the $L^{\infty}$ norm of the 
gradient of $A$ on the time interval 
$]T/2,T]$ which is independent of $\theta$. Similarly we can control
recursively higher order derivatives by writing down integral equations 
starting at various increasing times between $T/2$ and $T$.

In other words, even if we
started at time zero with a number $\theta$ which was quite large, after
a time of order at most one we get an estimate where the function and all
its derivatives up to order $k$ are of order unity.
 The result for all times then
follows immediately from previous results\ref{2}.
 Finally the result on the time
derivatives follows at once from the result on the space derivatives
using equation (A.1).\qed

We now need to prove that when $\sigma$ is small, if at the initial time
$A$ is of order unity, then it will decrease. Note that if
$\alpha\,=\,\beta\,=\,0$ this follows easily from the maximum principle
applied to the square of the modulus of $A$. In the general case, we
will use a method of local energy estimate\ref{3,2}.

\proclaim{Lemma A.2}. {Given the numbers $\alpha$, $\beta$, $\rho$, $S$,
and
$K$ as before, there is a number $\Theta>0$ and a number $\sigma_{0}>0$
such that if $\sigma\in[0,\sigma_{0}]$ and $t>\Theta$, we have
$$
\|A(t,\,\cdot\,)\|_{L^{\infty}}\le S^{-1}/2\;.
$$}

Note that we could get a bound of order $\sqrt\sigma$ but after some
larger time which depends on $\sigma$. One can also derive similar
estimates for the derivatives.

\proof The proof is essentially based on a local energy estimate\ref{2}. 
First of all we can get rid of the term $i\gamma\sigma A$ by considering
the function $e^{-i\gamma\sigma t}A(t,x)$.  We
introduce then a cut-off function depending on a positive parameter
$\epsilon$ 
$$
\varphi(x)={\epsilon^{d}\over(1+\epsilon^{2}|x|^{2})^{d}}
$$
where $d=1$ or $2$ is the dimension. One then introduces
the function of time
$$
\Psi(t)=\int\varphi(x)|A(t,x)|^{2}\;dx\;.
$$
It is easy to verify using integration by parts that since $A$ is
bounded,  this function satisfies
$$
{d\over dt}\Psi(t)\le-2\int\varphi(x)|\nabla A(t,x)|^{2}
+2\sigma\rho\int\varphi(x)|A(t,x)|^{2}
$$
$$
+2(1+|\alpha|)\int|\nabla\varphi(x)|
|\nabla A(t,x)||A(t,x|-2\int\varphi(x)|A(t,x)|^{4}\;.
$$
Completing the squares and using that $|\nabla
\phi|/\phi\le\Oun\epsilon$ we get for some number $a>1$ independent of
$\epsilon$ 
$$
{d\over dt}\Psi(t)\le
+2(\sigma\rho+a\epsilon^{2})\int\varphi(x)|A(t,x)|^{2}
-2\int\varphi(x)|A(t,x)|^{4}\;.
$$
Using the last term on the right hand side to control the first one, we
finally get
$$
{d\over dt}\Psi(t)\le
-(\sigma\rho+a\epsilon^{2})
\int\varphi(x)|A(t,x)|^{2}
+\Oun(\sigma\rho+a\epsilon^{2})^{2}\;.
$$
This implies since $\Psi(0)=\Oun$ that beyond a time
of order $(\sigma\rho+a\epsilon^{2})^{-1}$ we have
$$
\Psi(t)\le\Oun(\sigma\rho+a\epsilon^{2})\;.
$$
Let now $t$ be larger than the number $U$ of the previous Lemma. Then we
have 
$$
\|\nabla A(t,\,\cdot\,)\|_{L^{\infty}}\le \Oun\;.
$$
 From the previous
estimate we also know that the average of $|A|^{2}$ on any square
(segment) of size $\epsilon^{-1}$ is bounded by
$\Oun(\sigma\rho+a\epsilon^{2})$. Given $S$, this implies that in
dimension one we can
choose $\sigma_{0}$ and $\epsilon$ small enough such that the estimate holds.  
In dimension 2, one needs a similar energy estimate for the gradient to
apply an adequate Sobolev inequality. We
refer to the literature for analogous estimates. \qed
\beginsection{Appendix B: Proof of Lemmas II.4 and II.5.}

In this appendix we will always assume that the parameter $\eta$ is
positive and smaller than one.
We will first give an integral representation of the kernels of the
operators $L^t$ and $L^tR_{\delta,\theta}$. These operators
 are convolution operators with a sequence and we will 
estimate the $l^1$ norm of this sequence. The operator $L^t$ is the
convolution with the sequence $\und{\cal P}^t$ given by
$$
{\cal P}^t_n=(2\pi)^{-1}\int_0^{2\pi}e^{-in\varphi}(\eta+l(\varphi))^t
d\varphi 
\eqno (B1) 
$$
and the operator $L^tR_{\delta,\theta}$ is the
convolution with the sequence $\und{\cal M}^{t}$
$$
{\cal M}^t_n=(2\pi)^{-1}\int_0^{2\pi}e^{-in\varphi}(\eta+l(\varphi))^t 
\left(1-\psi\left({\omega-\varphi\over \delta\theta}\right)-
\psi\left({\omega+\varphi\over \delta\theta}\right)\right)
d\varphi\eqno (B2)
$$
since $2\pi/\delta\theta$ is larger than twice the diameter of the
support of $\psi$.

 First of all, since $l$ is $C^2$, we can find a positive
number $a<\min(\omega,\pi-\omega)$
 such that on the intervals $[\pm\omega-a,\pm\omega+a]$ we
have $|l''(\varphi)+D|<D/2$. Moreover, there is a positive constant $B<1$
such that
$$
\sup_{\varphi\notin]\pm\omega-a,\pm\omega+a[}|l(\varphi)|\le B\;.
$$

In the integral (B1) we separate the domain of integration in the union of
the two intervals $[\pm\omega-a,\pm\omega+a]$ and their complements.
Using integration by parts, one checks easily that the contribution
 of the complement to the integral is bounded
in $l^{1}$ norm  by $\Oun (1+t)^{2}(\eta+B)^t$.
 For the contribution of each interval we obtain by standard arguments\ref{9} 
$$
\left|(2\pi)^{-1}\int_{\pm\omega-a}^{\pm\omega+a}
e^{-in\varphi}(\eta+l(\varphi))^t d\varphi\right|\le \Oun e^{t\eta}t^{-1/2}\;.
$$
On the other hand, integrating twice by parts we get for $t\ge 2$ and
$n\neq0$
$$
n^2{\cal
P}^t_n=-(2\pi)^{-1}\int_0^{2\pi}e^{-in\varphi} 
\left(tl''(\varphi)(\eta+l(\varphi))+t(t-1){l'(\varphi)}^2\right)
(\eta+l(\varphi))^{t-2}d\varphi 
$$
and by similar estimates as before, for $n\neq 0$ and $t\ge 2$
$$
|{\cal P}^t_n|\le \Oun n^{-2}t^{1/2}e^{\eta t}\;.
$$
Combining the two estimates we get
$$
|{\cal P}^t_n|\le \Oun (1+n^2/t)^{-1}t^{-1/2}e^{\eta t}\;.
$$
which implies immediately
$$
\|{\cal P}^t\|_{l^1}\le \Oun e^{\eta t}\;,
$$
and this proves Lemma II.4.

We now come to the proof of Lemma II.5. First of all, if
$t<(\delta\theta)^{-2}$, the estimate follows at once from Lemma II.4 and
Lemma II.1 (as long as $\delta^2\theta^2>\eta$ which is always
assumed). From now on we will assume $t\ge (\delta\theta)^{-2}$.

It follows by estimates similar to the previous ones that 
$$
|{\cal M}^t_n|\le \Oun
(1+n^2/t)^{-1}t^{1/2}e^{-t(D\delta^2\theta^2/2-\eta)}\;. 
$$
from which Lemma II.5 follows.

\beginsection{Appendix C. Estimation of oscillating sums.}

\def\cald{{\cal D}}
\proclaim{Lemma C.1}. {There is a constant $C_{9}>0$, a constant
$a>0$ and a constant $1>\eta_{0}>0$ such that if $\eta\in[0,\eta_{0}]$, 
if $\delta$, $\theta$, $B$ and $A$ satisfy the hypothesis of
Lemma II.3, if the sequence  $\und f^{t}$ is defined by
$$
f^{t}_{n}=A^{3}(\delta^{2}t,n\delta\sqrt{2/D})\;,
$$ 
then for any integers $s$ and $t$ we have
$$
\|L^{s} \underline{e^{(3i\,\cdot\,)}f^{t}}
\|_{l^{\infty}}\le C_{9}\left(\delta\theta e^{\eta s}+
e^{-as}\right)\;.
$$}

\proof From hypothesis {\bf H3} it follows
 that there is a number $\Upsilon>0$
such that $|4\omega \pmod {2\pi}|>\Upsilon$
 and $|2\omega\pmod {2\pi}|>\Upsilon$. We can
now write for the kernel of the operator $L^{s}$
$$
L^{s}(n,m)={1\over 2\pi}\int e^{i\phi(n-m)}(\eta+l(\phi))^{s}d\phi
={1\over 2\pi}\int_{|\phi\pm\omega|>\Upsilon}
 e^{i\phi(n-m)}(\eta+l(\phi))^{s}d\phi
$$
$$
+{1\over 2\pi}\int_{|\phi-\omega|\le\Upsilon}
 e^{i\phi(n-m)}(\eta+l(\phi))^{s}d\phi
+{1\over 2\pi}\int_{|\phi+\omega|<\Upsilon}
 e^{i\phi(n-m)}(\eta+l(\phi))^{s}d\phi\;.
$$
Using techniques analogous to those of the previous appendix, it is easy
to show that the first integral is the kernel of an operator with norm
$\Oun e^{-as}$ in $l^{\infty}$ for some constant $a>0$ which can be
chosen independent of $\eta$ if $\eta_{0}>0$ is small enough.
 The second and third integrals are estimated by 
similar methods of ``summation by parts'' 
and we will only treat the second one.

We  observe that
$$
\sum_{m}f_{m}^{t}{1\over 2\pi}\int_{|\phi-\omega|\le\Upsilon}
 e^{i\phi(n-m)+3i\omega m}(\eta+l(\phi))^{s}d\phi
$$
$$
=\sum_{m}(f_{m}^{t}-f_{m+1}^{t})
{1\over 2\pi}\int_{|\phi-\omega|\le\Upsilon}
 {e^{i\phi(n-m)+3i\omega m}(\eta+l(\phi))^{s}\over 
1-e^{i(\phi-3\omega)}}d\phi\;.
$$
From the properties of the function $A$ (see Lemma A.1) we have
$$
|f_{m}^{t}-f_{m+1}^{t}|\le\Oun\delta\theta\;.
$$
Using again stationary phase methods, it follows as in the previous
section that 
$$
\left|{1\over 2\pi}\int_{|\phi-\omega|\le\Upsilon}
 {e^{i\phi(n-m)+3i\omega m}(\eta+l(\phi))^{s}\over 
1-e^{-i(\phi-3\omega)}}d\phi\right|\le{\Oun s^{-1/2}e^{\eta s}\over
1+(n-m)^{2}/s}\;, 
$$
and the Lemma follows.
\qed
\proclaim{Lemma C.2}. {Let $(q_{n,m})$ be such that 
$$
\sum_{n,m}(1+|n|+|m|)^{k+2}|q_{n,m}|<\infty\;,
$$
and $q(\phi_{1},\phi_{2})=0$ on a neighborhood of the four points 
$(\pm\omega,\pm\omega)$.
Then there is a number $\delta_{0}>0$ such that if
 $\und\cala$ is constructed  as in III.2 with an $A(x)$ such
that there is a constant $M$ and a number $k>0$ such that
$$
\sup_{0\le j\le k}\theta^{_j}\|\partial^{j}A\|\le M\;,
$$
then
$$
\sup_{r}\left|\sum_{n,m}q_{r-n,r-m}\cala_{n}\cala_{m}\right|\le
\Oun\delta^{k}\theta^{k}
$$}

\proof The proof again essentially mimics the proof of the stationary phase
estimates, namely it relies on an integration by parts.
Let $F(x)=(3g)^{-1/2}A(\delta^{2}t,\sqrt{2/D} (x+\delta Vt))$,
 we have to consider four terms which are
essentially of the same form. We will only treat one of  them in
details, the others are estimated similarly. We  have
$$
\sum_{n,m}q_{r-n,r-m}e^{i(\omega (n+m))}F(\delta n)F(\delta m)=
$$
$$
e^{2i\omega r}
\sum_{n,m}F(\delta n)F(\delta m) {1\over4\pi^{2}}\int
q(\phi_{1},\phi_{2}) e^{i((\phi_{1}-\omega)(r-n)
+(\phi_{2}-\omega)(r-m))} d\phi_{1}, d\phi_{2}\;.
$$
Let $\varpi$ be a positive number such that $q(\phi_{1},\phi_{2})=0$ on
the four sets $[\epsilon_{1}\omega-\varpi,\epsilon_{1}\omega+\varpi]\times
[\epsilon_{2}\omega-\varpi,\epsilon_{2}\omega+\varpi]$ where
$\epsilon_{1}=\pm1$ and $\epsilon_{2}=\pm1$ . We now decompose the integration
domain into a finite union of domains, in each of which $\phi_{1}$ or
$\phi_{2}$ (but may-be not both) is at a distance at least $\varpi$ of
$\pm\omega$. Let us call $\cald$ such a domain, and assume for
definiteness that on $\cald$ we have 
$$
\inf_{(\phi_{1},\phi_{2})\in\cald}|\phi_{1}\pm\omega|>\varpi\;.
$$
It is then easy to verify that
$$
\sum_{n,m}F(\delta n)F(\delta m) {1\over4\pi^{2}}\int_{\cald}
q(\phi_{1},\phi_{2}) e^{i((\phi_{1}-\omega)(r-n)
+(\phi_{2}-\omega)(r-m))} d\phi_{1} d\phi_{2}=
$$
$$
\sum_{n,m}F(\delta (r-n))F(\delta (r-m)) {1\over4\pi^{2}}\int_{\cald}
q(\phi_{1},\phi_{2}) e^{i((\phi_{1}-\omega)n
+(\phi_{2}-\omega)m)} d\phi_{1} d\phi_{2}=
$$
$$
\sum_{n,m}(F(\delta (r-n))-F(\delta (r-n+1))
F(\delta m) {1\over4\pi^{2}}\int_{\cald}
{q(\phi_{1},\phi_{2})\over 1-e^{i(\phi_{1}-\omega)}}
 e^{i((\phi_{1}-\omega)n+(\phi_{2}-\omega)m)} d\phi_{1}, d\phi_{2}\;.
$$
From our hypothesis, we have
$$
|F(\delta (r-n))-F(\delta (r-n+1))|\le\Oun \delta\theta\;.
$$
The result follows by several applications of the argument, and
controlling the convergent of the ensuing sum by the summability
properties of $(q_{n,m})$.\qed

\proclaim{Lemma C.3}. { There is a number $b>0$ such that if
$0\le\eta<1$, if the sequence $(q_{n,m)}$ is such that
$$
\sum_{n,m}(1+|n|+|m|)^{2}|q_{n,m}|<\infty\;,
$$
and $q(\phi_{1},\phi_{2})=0$ on a neighborhood of
$\phi_{1}+\phi_{2}=\pm\omega$,  then for any $t\ge 0$ we have
$$\|\call^{t}\und q(\und v\,\und w)\|_{l^\infty}\le
\Oun e^{-bt}\|\und v\|_{l^\infty}\|\und w\|_{l^\infty}\;.
$$}

\proof We have easily the formula
$$
\big(\call^{t}\und q(\und v\,\und w)\big)_{r}\hskip -1.5 pt=
\hskip -1.5 pt\sum_{n,p}v_{n}w_{p}
{1\over4\pi^{2}}\int (l(\phi_{1}+\phi_{2})+\eta m(\phi_{1}+\phi_{2}))^{t}
q(\phi_{1},\phi_{2})e^{i(\phi_{1}(r-n)+\phi_{2}(r-p))}d\phi_{1}d\phi_{2}\;. 
$$
The result follows at once from the fact that on the support of $q$ we have
$\phi_{1}+\phi_{2}\neq\pm\omega$. The necessary summability results are
obtained by integration by parts as in the previous Lemma.\qed
\bigskip

\noindent {\sl Acknowledgments}. This paper was inspired and 
initiated during the Marseille workshop "Dynamics, Stochastics 
and Complexity". The author is very grateful to R.Lima for his kind
invitation to this meeting and for fruitful discussions. 

\beginsection{References.}

\item{1.} J.Bricmont, A.Kupiainen. Renormalizing Partial
Differential Equations,  in {\sl XIth International
congress on mathematical physics}, D.Iagolnitzer ed. International Press
Incorporated, Boston (1995).\medskip
\item{2.}P.Collet.Thermodynamic limit of the Ginzburg-Landau equation.
Nonlinearity {\bf 7}, 1175-1190 (1994).\medskip
\item{3.}P.Collet, J.-P.Eckmann. The time-dependent amplitude
equation for the Swift-Hohen\-berg problem. Commun. Math. Phys. {\bf 132},
139-153 (1990).\medskip
\item{4.} L-Y.Chen, N.Goldenfeld, Y.Oono. Renormalization group
theory for global asymptotic analysis. Phys. Rev. Lett. {\bf 73},
1311-1315 (1994). \medskip
\item{5.} M.C.Cross, P.C.Hohenberg.  Pattern formation outside of
equilibrium. Rev. Mod. Phys. {\bf 65}, 851-1113 (1993).\medskip
\item{6.} W.Eckhaus. The Ginzburg-Landau manifold is an attractor. J.
Nonlinear Sci. {\bf 3}, 329-348 (1993). \medskip
\item{7.} N.Goldenfeld. {\sl  Lectures on phase transitions and the
renormalization group}. Addison-Wesley, Reading (1992).
See also N.Goldenfeld et al., Anomalous dimensions and the
renormalization group in a nonlinear diffusion process. 
Phys. Rev. Lett. {\bf 64}, 1361-1364 (1990).\medskip
\item{8.} G.Guckenheimer, P.Holmes. {\sl Nonlinear Oscillations,
Dynamical Systems, and bifurcations of Vector Fields.} Springer Verlag, 
 Berlin Heidelberg New York 1983.\medskip
\item{9.} L.Hormander. {\sl The analysis of linear partial
differential operators I : distribution theory and Fourier analysis}.
Springer Verlag, Berlin Heidelberg New-York (1990).\medskip
\item{10.} Y.Il'yashenko, S.Yakovenko. Smooth normal forms of local
families of diffeomorphisms and vector fields. Russian Math. Survey {\bf
46}, 1-43 (1991).\medskip
\item{11.}P.Kirmann, G.Schneider, A.Mielke.  The validity of
modulation equations for extended systems with cubic nonlinearities.
Proc. Roy. Soc. Ed. {\bf A 122}, 85-91 (1992).\medskip
\item{12.} J.Lega, A.Newell, T.Passot. Order parameter equations
for patterns. Ann. Rev. Fluid. Mech. {\bf 25}, 399-453 (1993).\medskip
\item{13.} G.Schneider. Global existence via Ginzburg-Landau formalism
and pseudo orbits of Ginzburg-Landau approximations. Comm. Math. Phys.
{\bf 164}, 159-179 (1994).\medskip
\item{14.}G.Schneider. Analyticity of Ginzburg-Landau modes. J. Diff.
Equ. {\bf 121}, 233-257 (1995).\medskip
\item{15.}A. Van Harten. On the validity of Ginzburg-Landau's
equation. J. Nonlin. Sci. {\bf 1}, 397-422 (1991).\medskip
\bye